\documentstyle[proceedings,numreferences,epsf]{crckapb}
\begin{opening}
\title{Observation of scaling behavior 
in a Coulomb blockade system}
\author{Karsten Flensberg}
\institute{Dansk Institut for Fundamental Metrologi, \\
Bygn. 307, Anker Engelunds Vej 1\\ DK-2800 Lyngby, Denmark}
\author{L.W. Molenkamp}
\institute{2. Physikalisches Institut, RWTH Aachen\\ D-52056 Aachen, Germany}
\end{opening}
\runningtitle{Scaling in Coulomb Blockade}
\begin{document}

\begin{abstract}  
We describe two experiments to study the influence of fluctuations in the 
electron charge on the transport properties of a quantum dot. First, we
scan a device from single- to double quantum-dot behavior by varying the
conductance of a connecting point contact. Second, we measure the dependence
of the charging energy on the conductance of the barriers.
The experiments are compared with traces obtained from a theory 
based on a Luttinger liquid type description. This theory predicts a
scaling behaviour of the charging energy, in good agreement with
our experiments. 
\end{abstract}
%PACS numbers: 73.40.Gk, 73.50.Dn, 73.60.B\\
%Keywords: Quantum-dots, Coulomb-blockade, (Al,Ga)As

\section{Introduction}
\label{sec:intro}

During the past few years there has been considerable 
interest in the transport properties of quantum dots\cite{likh:rev}. 
A quantum dot consists of a small metallic island,
coupled to external leads by tunnel barriers. 
At sufficiently low temperatures the energy associated with the addition 
of one electron to the dot (i.e. the charging energy
$ U \equiv e^2 / C$, with $C$ the total dot capacitance) 
exceeds the thermal energy ($k_{\rm B}T$) and the
conductance of the dot is suppressed. 
This is the Coulomb blockade of the conductance. 
It can be overcome by adjusting the electrostatic potential of the dot,
which is conveniently achieved by varying the voltage $V_{\rm g}$ 
of an external gate electrode.
At certain values of $V_{\rm g}$ the electrostatic energy of the dot 
containing $n$ electrons becomes equal to that for
$(n+1)$ electrons. At these gate 
voltages transport becomes possible. This results in a periodic series of
conductance peaks  as a function of $V_{\rm g}$, known as Coulomb 
oscillations. 

The Coulomb oscillations can only be observed when the electrons
are tightly confined to the quantum dot, 
i.e. when the conductances of the barriers 
connecting the dot to external leads are small. When the barrier
height is decreased, the charge on the island becomes less well-defined due
to quantum charge fluctuations and the Coulomb blockade is smeared. 
This has been studied both
experimentally\cite{metal:exp} and theoretically\cite{metal:theory}
in metalic systems which are characterized by a
large number of channels connecting the island to the leads. 
In these systems the Coulomb blockade is gradually reduced as the
conductance is increased above the conductance quantum, $2e^2/h$.

For electrostatically defined dots in AlGaAs heterostructures, 
the contacts to the leads are quantum point contacts with a 
small number of channels. In this case a fully opened contact has a 
conductance larger than $2e^2/h$ and the Coulomb blockade is 
found to disappear\cite{kouw91,foxm93}.  
This phenomenon has attracted much theoretical attention, 
since it proved possible to described this behaviour in terms of
models know from one-dimensional interacting electron 
gases\cite{flen93capa,flen94capa,matv95,furu95},
so-called  Luttinger liquids. The studies of these non-Fermi liquid
many electron systems has recently gained new impetus because of its
possible relevance to mesoscopic semiconductor structures, both in 
describing electron tunneling behavior\cite{kane92fish,matv93glaz} 
and at low magnetic fields\cite{alei95glaz2}, as well as 
in quantum Hall systems where the
edge states has been suggested to be describable in terms of
Luttinger liquids\cite{wen90}. An interesting aspect of these
suggestions is the possibility of observing effects such as
power-law behaviors know from 
one-dimensional physics in systems which are not truly
one-dimensional.
 
The quantum dot connected to the leads by quantum point contacts may be
mapped to a system of coupled one dimensional electron gases,
Luttinger liquids\cite{flen93capa,flen94capa,matv95}, and it can be
argued that that the charging energy associated
with the addition of an electron to the dot should scale
with the conductance of the point-contact barriers, according 
to\cite{flen93capa,flen94capa}  
\begin{equation}
U^* \sim U ( 1 - T )^{N_c}.
\label{scafor}
\end{equation}
Here, $U$ is the bare charging energy and $U^*$ is the effective 
(or "renormalized") charging energy observed for finite barrier conductance, 
$N_c$ the total number of quantum point contacts leading to the dot, and 
$T$ the transmission probability of each contact. Here the
transmission coefficients are assumed to be equal. For a situation
with  different $T$'s the scaling will be governed by the larger of
the transmission coefficients.

In this paper we describe two experiments on a double quantum dot
device that were designed to verify the validity of Ref.\ \ref{scafor}.
First, we study the transition from single- to  double-dot
behaviour as a function of the transmission of the barrier
seperating the two dots. Second, in the double dot regime,
we use one dot as an
electrometer that measures the charging energy of the other dot. 
Both experiments can be well modeled by using a standard rate
equation approach for the coupled dot system, but with the important
addition that we allow the effective energy gap to be modified
according to Eq.\ (\ref{scafor}).
Part of this research was previously reported by us in ref. \cite{mole95}. 
Here we enlarge the discussion and also present further 
experimental evidence for the interpretation in terms of the scaling law
of Eq.\ (\ref{scafor}).  
Recently, also Waugh {\it et al.}\cite{waug95} reported an experiment
which has been shown to be reflect one-dimensional 
physics\cite{matv:pre,gold:pre}.

The paper is organized as follows. In Sec. \ref{sec:setup} we 
describe the experimental realization and the operation principle
of the coupled dot system, in Sec. \ref{sec:model} the theoretical model
is derived and the experimental results and the comparisons with the
model are given in Secs. \ref{sec:flucn2exp} and \ref{sec:flucndiff}.

\section{Experimental setup and the electrometer operation}
\label{sec:setup}

\begin{figure}
\vbox to 8.0cm {\vss\hbox to 8cm
 {\hss\
   {\includegraphics{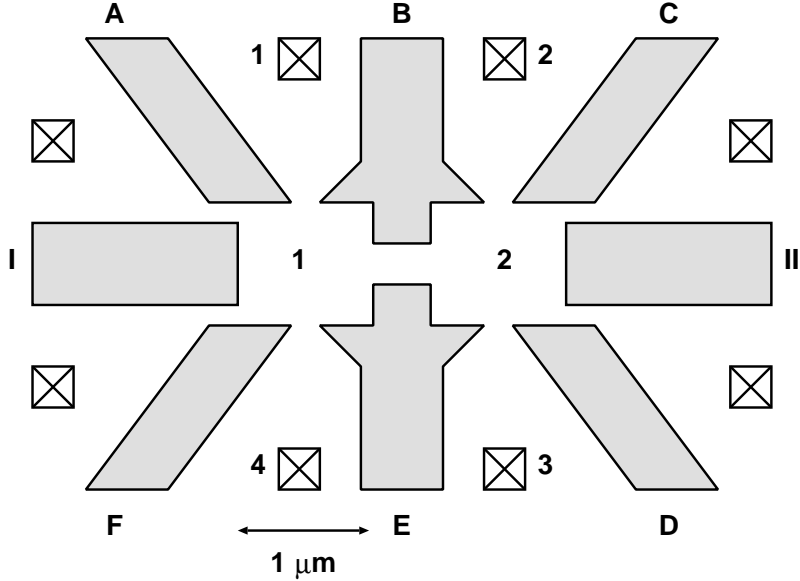}
   }
  \hss}
}
\caption{
Schematic lay-out of the double dot-sample. The hatched areas denote
gates, the crosses ohmic contacts. The lithographic width of one dot is
1 $\mu$m, the distance between two gates defining a point contact is
typically 250 nm. We denote the conductance of the contacts
connecting to dot 1 by 
$g_1=G_{\rm{AB}}$ and $g_2=G_{\rm{EF}}$, the
contacts of dot 2 by $g_3=G_{\rm{BC}}$ and $g_3=G_{\rm{ED}}$, the
contact connecting the two dots by $g_5=G_{\rm{EB}}$, and the
voltage of the four terminals as $V_i$, $i=1,\ldots,4$, corresponding
to the terminal numbers shown in the figure.}
\label{fig:layout} 
\end{figure}
\begin{figure}
\vbox to 10.5cm {\vss\hbox to 10.5cm
 {\hss\
   {\includegraphics{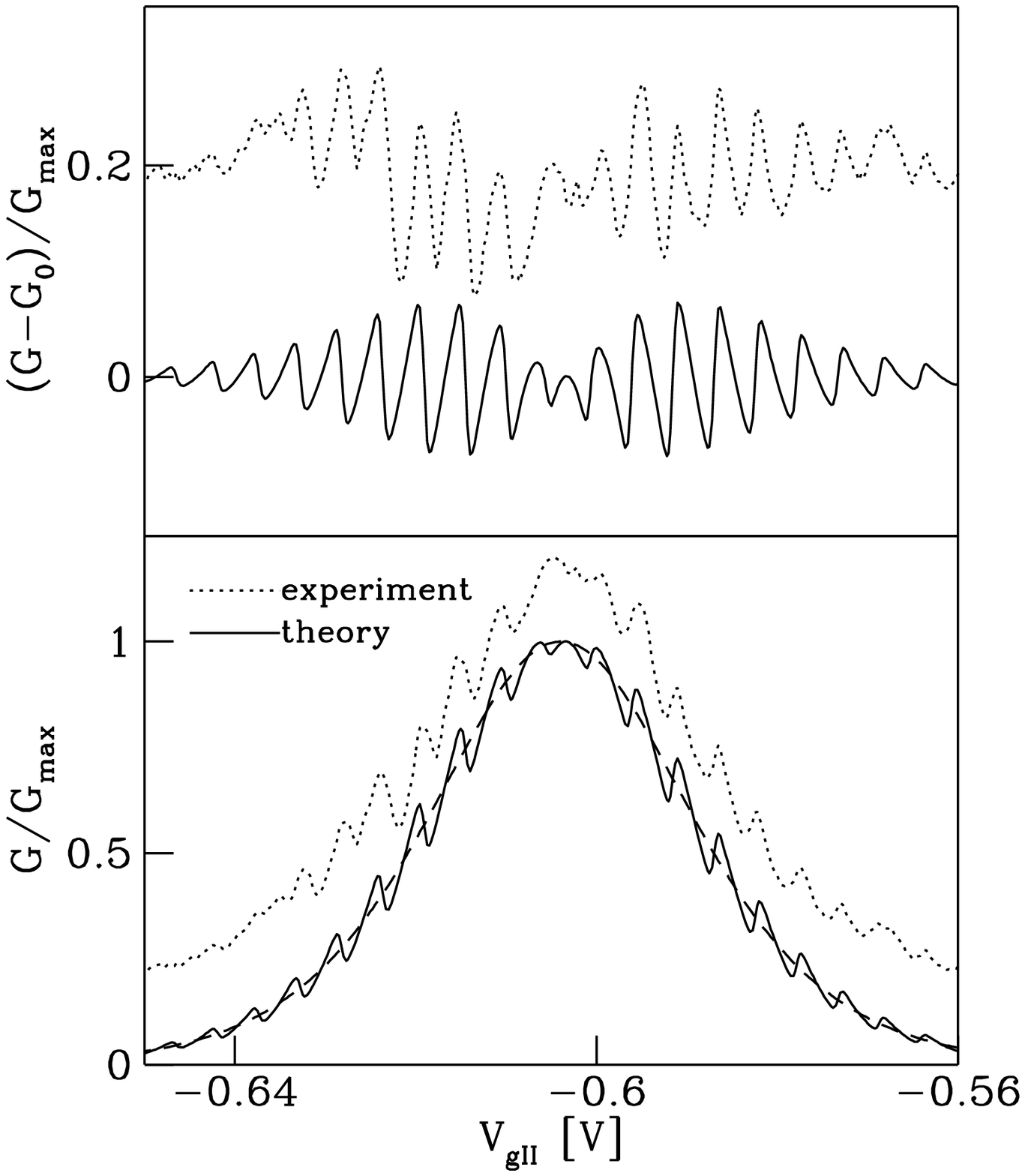}
   }
  \hss}
}
\caption{The operating principle of the electrometer experiment:
scanning $V_{\rm{gII}}$ leads to an increase in
$n_2$, the number of electrons on dot 2, which gives sawtooth
oscillations in the potential on dot 1 and hence to the conductance
through dot 1.
Full lines are results of the model calculation and the bottom
panel shows a fit to the experimental trace (dotted curve).
The dashed line corresponds to the case where there is no
charging energy for the second dot. In the middle panel
we have subtracted the conductance of the first dot from the result 
obtained when there is no charging energy of the second dot. 
The size of the resulting oscillations
thus measures the effective charging energy of dot 2, which is 
used for determining $U_2^*$, as discussed below.}
\label{fig:elmeter} 
\end{figure}

The devices used in this study are fabricated from a (Al,Ga)As modulation doped
heterostructure that contains a two-dimensional electron gas (2DEG) of mobility
$2 \times 10^6$ cm$^2$V$^{-1}$s$^{-1}$ and sheet carrier concentration 
$1.8 \times 10^{11}$ cm$^{-2}$. The nanostructures are defined
electrostatically by split gates, 
structured using both optical and electron-beam
lithography. A schematic lay-out of the device is shown in Fig.\
\ref{fig:layout}. 
The diameter of each dot is about 1 $\mu$m. 
All experiments are performed in the mixing chamber of a dilution
refrigerator with a minimum base temperature of 20 mK (we estimate
the eletron gas temperature to be 150 mK).

The device is designed to give a very large
control over the device parameters since the conductance of all contacts
can be adjusted. This means that conductance traces of the quantum
dot can be taken for
different values of the conductances of the relevant point contacts
barriers. Inevitably, there is some cross-talk between different
gates which  makes the determination of the contact resistances uncertain.
We have included this effect in the error estimate of the first
experiment. For the second
experiment where we study the transition from one to two dots, the
cross-talk give rise to an off-set that we determine by fitting to
our model. In the experimental results that will be presented, we 
have measured the conductance $G \equiv G_{14}$ 
between ohmic contacts 1 and 4, using standard low-frequency 
lock-in techniques. 

Our device can be operated as a semiconductor analogue of the metallic 
two-dot electrometer device described by Lafarge et al.\cite{lafa91}. 
The mode-of-operation of the electrometer experiment is
schematically depicted in the bottom panel of Fig.\ \ref{fig:elmeter}. 
What is measured is the dependence of $G$ on $V_{\rm{gII}}$, 
the voltage on gate II. 
In our double-dot device, scanning gate II stepwise increases of 
the occupancy in both dot 1 and dot 2, but with a much shorter
stepsize (in  $V_{\rm{gII}}$) in dot 2, 
because of the larger dot-to-gate capacitance.
The trace in  Fig.\ \ref{fig:elmeter} shows the conductance within a 
$V_{\rm{gII}}$-voltage range where $n_1$ only changes by 1, whereas
each superimposed sawtooth corresponds to the addition of
one electron to dot 2. 
It is then clear that the size of the sawteeth is a measure of the 
charging energy of dot 2; dot 1 thus acts as an 
electrometer\cite{lafa91} that measures the changes in the potential of dot 2. 
In the absence of a charging energy of dot 2, the conductance 
is given by usual Coloumb oscillation peaks, which we have
indicated with the dashed line (obtained by setting $U_2^*=0$ in the
equations below). 
The full curve in Fig.\ \ref{fig:elmeter} is, in fact, a fit
to an experimental trace (dotted curve), using a theory we will
discuss below. 

Subtraction of the dashed and the 
full line gives the trace shown in the top panel of the figure.
We have also subtracted the same background from the 
experimental trace. 
Using the subtracted trace, we are able to fit the experimental
curves and this fitting procedure is used to determine an effective
charging energy for dot in Sec. \ref{sec:fitting}.

\section{Setting up the model}
\label{sec:model}

In this section, we describe the theoretical modelling that we use.
We apply the standard electrostatic model for the interaction
energy and use a rate equation approach to solve for the conductance 
of the device. Then we incorporate the charge fluctuations in our
model by introducing a renormalized charging energy.

\subsection{The electrostatic energy}
\label{sec:electro}

The electrostatic energy of our double dot system can be found by
considering the various capacitances involved. We define 
$C_i$ as the capacitance between dot $i$ and the
leads, $C_{12}$ as the capacitance between the two dots,
$C_g$ describes the capacitance between a dot and its
nearest gate, while $C_g'$ is the capacitance to the
opposite gate. The device is assumed to be symmetric so
that  $C_{g1}=C_{g2}$ and $C_{g1}'=C_{g2}'$.
Following, e.g., Ruzin {\em et al.}\cite{ruzi92} the  electrostatic
energy is then found to be,
\begin{eqnarray}
E(n_1,n_2) &=& U_{11} n_1^2 + U_{22} n_2^2 + U_{12}n_1n_2 
\nonumber\\
&&+(a_{11} eV_{g1} +a_{12} eV_{g2})n_1+(a_{21} eV_{g1} +a_{22} eV_{g2})n_2,
\end{eqnarray}
where $n_i$ is the number of electrons on dot $i$, and where
\begin{eqnarray}
U_{11} &=& \frac{2C_2+C_g+C_g'+C_{12}}{D},\nonumber\\
U_{12} &=& \frac{C_{12}}{D},\\
a_{11} &=& \frac{C_gC_g'+[C_g']^2+C_{12}(C_g+C_g')+2C_g'C_2}{D},\nonumber\\
a_{12}&=& \frac{C_g^2+C_gC_g'+C_gC_{12}+2C_gC_2}{D},\nonumber\\
D &=& [C_g+C_g']^2+2(C_1+C_2)(C_g+C_g'+C_{12})+4C_2C_1\nonumber\\
&& +2C_{12}(C_g+C_g'),\nonumber
\end{eqnarray}
and the quantities $U_{22}$, $a_{22}$, and $a_{21}$ are obtained by
interchanging  indices accordingly. 

From fitting our experimental data (in absence of quantum
fluctuations), we obtain
\begin{eqnarray}
U_{ii} &\approx& 0.13 \,\mbox{meV}\nonumber\\
U_{12} &\approx& 0.009\,\mbox{meV}\nonumber\\
a_{ii} &\approx& -0.20\nonumber\\
a_{12}&\approx& -3.12\nonumber
\end{eqnarray}

The electrostatic energy is minimized when
\begin{equation}
\left(
\begin{array}{l}
n_1 \\
n_2
\end{array}\right)=
\left(\begin{array}{rr}
2 U_1 & U_{12}\\
U_{12} & 2 U_2
\end{array}\right)^{-1}
\cdot \left(\begin{array}{rr}
a_{11} & a_{12}\\
a_{21} & a_{22}
\end{array}\right)
\left(\begin{array}{c}
V_{g1}\\
V_{g2}
\end{array}\right)
\equiv
\left(\begin{array}{l}
n_{10} \\
n_{20}
\end{array}\right).
\end{equation}
At zero temperature the number of electrons on the dots is therefore given by
\begin{equation}
\langle n_i \rangle = \mbox{\rm Int} ( n_{i0}).
\end{equation}
Denoting the deviation from the number of electrons that minimize
the electrostatic energy by $\delta n_i = n_i - n_{i0}$, we get
\begin{equation}\label{Edelta}
E  = U_1 \delta n_1^2+ U_2 \delta n_2^2 + U_{12} \delta n_1 \delta n_2
+ E(n_{10},n_{20}).
\end{equation}

\subsection{Solution of the rate equation}
\label{sec:rate}

Next we consider the rate equations for the device. The analysis
follows that of Kulik and Shehkter\cite{kuli75}, Glazman and
Shehkter\cite{glaz89}, or Beenakker\cite{been91}.
The layout is shown in Fig.\ \ref{fig:layout} and we denote
the conductances and voltages as explained in the figure and the
probability of having charge configuration $(n_1,n_2)$ in the two
dots as $W(n_1,n_2)$. We set up the condition for 
steady state for dot 1 as
\begin{eqnarray}
&&\frac{d W(n_1,n_2)}{dt}=\nonumber\\
&&-W(n_1,n_2)\sum_{i=1,4}g_i\sum_\pm
f(E(n_1,n_2)-E(n_1\pm 1,n_2)\pm eV_1)\nonumber\\
&&-W(n_1,n_2)\sum_{i=2,3}g_i\sum_\pm
f(E(n_1,n_2)-E(n_1,n_2\pm 1)\pm eV_1)\nonumber\\
&&+\sum_\pm W(n_1 \pm 1,n_2) \sum_{i=1,4}
g_if(E(n_1\pm 1,n_2)-E(n_1,n_2)\mp eV_i)\nonumber\\
&&+\sum_\pm W(n_1,n_2\pm 1) \sum_{i=2,3}
g_i f(E(n_1,n_2\pm 1)-E(n_1,n_2)\mp eV_i)\nonumber\\
&&-W(n_1,n_2) g_5\big[f(E(n_1,n_2)-E(n_1+1,n_2-1))\nonumber\\
&&+f(E(n_1,n_2)-E(n_1-1,n_2+1))\big]\nonumber\\
&&+W(n_1+1,n_2-1)g_5f(E(n_1+1,n_2-1)-E(n_1,n_2))\nonumber\\
&& +W(n_1-1,n_2+1)g_5f(E(n_1-1,n_2+1)-E(n_1,n_2))=0.
\end{eqnarray}
Here the function $f$ is the is the usual function given
by Fermis Golden Rule,
\begin{equation}
f(E) =\frac{E}{1-e^{-\beta E}}.
\end{equation}
In equilibrium the distrubution function $W$ is given by
\begin{equation}
W_0(n_1,n_2) = \exp(-\beta E(n_1,n_2))/\sum_{n_1,n_2}\exp(-\beta E(n_1,n_2)).
\end{equation}
In order to find the conductances, we expand in powers of $V_i$ and
we use the following relations
\begin{eqnarray}
&&W_0(x)\left[f(E(x)-E(y)+eV)-W_0(y)f(E(y)-E(x)-eV)\right]\nonumber
\\
&&\approx  eV\beta f(E(x)-E(y))W_0(x)=eV f(E(y)-E(x))W_0(y),\nonumber
\\
&& W(n_1,n_2) =  W_0(n_1,n_2)+W_1(n_1,n_2),\nonumber
\end{eqnarray}
where $W_1$ is of order $V_i$.

Since we have a special interest in the case when the device is biased
such that current is passed only through dot number 1 and
where $g_1=g_4=G_0$, we set $V_3=V_4=0$. Furthermore, we let the
bias across dot 1 be symmetric such that $V/2=V_1=-V_2$ (any
asymmetry can be absorbed in gate voltages). 
Under these conditions is it straightforward to verify that 
$W_1=0$ is a solution to the steady state criteria above. 

We are now in a position to calculate the current by from the
equation
\begin{eqnarray}
I_1 &=& \frac{g_1}{e}\sum_{n_1n_2} \left[
W(n_1,n_2)f(E(n_1,n_2)-E(n_1+1,n_2)+eV_1)\right.\nonumber\\
&& \left.-W(n_1+1,n_2)f(E(n_1+1,n_2)-E(n_1,n_2)-eV_1)\right],
\end{eqnarray}
and in the special case considered here we get
\begin{equation}\label{G}
G = \frac{G_0}{4kT}\sum_{n_1,n_2}W_0(n_1,n_2)f(E(n_1,n_2)-E(n_1-1,n_2))
\end{equation}

\subsection{Modelling of quantum charge fluctuations}
\label{sec:fluc}
       
Next, we address the question how quantum fluctuations modify the
transport properties of the dot device. Assuming an electrostatic
model for the interaction energy, it is clear from
Eq. (\ref{Edelta}) that the {\em period of the Coulomb oscillations
is uneffected by quantum fluctuations}. This is because if the
Hamiltonian is given by
\begin{equation}
H = H_0+E(n_1,n_2),
\end{equation}
(here $H_0$ is the single electron Hamiltonian)
it is not changed when changing $n_{i0}$ by one 
apart from adding a constant to the energy and changing the
corresponding average number of electrons by one.
Therefore, changing the gate voltage such that the average number of
electrons in one of the dots is increased by one does
not change the transport properties.
\footnote{This argument assumes that the density of states and the 
tunneling matrix elements are constant within the range of the
chemical potential that we are considering.} 
Thus, the spacing of the peak structures are determined by the
electrostatics of the system, and only the widths and amplitudes
of the peaks are altered by an increase of the number fluctuations.
We will model this by invoking a renormalized charging energy 
$\delta E^*$ as described in the following two subsections. 
This renormalized charging energy is then inserted into 
Eq.\ (\ref{G}) and we have a new expression for the conductance
which takes the quantum fluctuations into account
\begin{equation}\label{Grenor}
G = \frac{G_0}{4kT}\sum_{n_1,n_2}W_0^*(n_1,n_2)f(\delta E^*(n_1,n_2)-
\delta E^*(n_1-1,n_2)),
\end{equation}
where
\begin{equation}
W_0^*(n_1,n_2) = \exp(-\beta \delta E^*(n_1,n_2))/\sum_{n_1,n_2}
\exp(-\beta \delta E^*(n_1,n_2)).
\end{equation}

\subsubsection{Fluctuations of the charge on dot 2}
\label{sec:flucn2}

In one of our experiments the charge on dot 2 is
allowed to fluctuate (Sec. \ref{sec:flucn2exp}) by lowering 
the barriers BC and DE. The resulting increase
in variations of $n_2$ will be modelled by using a renormalized value
for the charging energy that controls the deviation away from the 
optimum number of electrons on dot 2. We have 
\begin{equation}\label{Erenor1}
\delta E^*(n_1,n_2)= [U_1-U_{12}^2/U_2]
\delta n_1^2+U_2^*(\delta n_2-\delta n_1 U_{12}/2U_2)^2.
\end{equation}

\subsubsection{Fluctuations of the charge difference between the dots}
\label{sec:flucndiff}

In a second experiment, we allow for the charge to fluctuate between 
the two dots by gradually opening contact $g_5$ defined between gates
B and E. This results in an effective smaller charging energy for 
the charge difference between the dots. This we now incorporate in
our modelling. Writing the bare electrostatic
energy in terms of this difference and the total number of
electrons on the two dots, $\Delta n = \delta n_1 - \delta n_2$
and  $N=n_1+n_2$, we have that
\begin{equation}
E(\Delta n,N) = U_N (N-N_0)^2+U_{\Delta n}(\Delta n-\Delta n_0)^2
\end{equation}
where $N_0 = n_{10}+n_{20}$, $\Delta n_0 = n_{10}-n_{20}$, and where
\begin{eqnarray}
U_N &=& (2 U_1+U_{12})/4\\
U_{\Delta n} &=& (2U_1-U_{12})/4.
\end{eqnarray}
Here we have assumed that $U_1=U_2$, which is a good approximation in
our experiment.

Now in the case where the charge difference $\Delta n$ exhibit
quantum fluctuations it may be modeled by renormalizing
the charging energy associated with changes in $\Delta n$.
This leads to the model for the renormalized energy
\begin{equation}\label{Erenor2}
\delta E^*(\Delta n,N) = U_N (N-N_0)^2+U_{\Delta n}^*(\Delta n-\Delta n_0)^2.
\end{equation}

\section{Fluctuations of the charge on dot 2, $N_c=2$}
\label{sec:flucn2exp}

\begin{figure}
\vbox to 12cm {\vss\hbox to 12cm
 {\hss\
   {\includegraphics{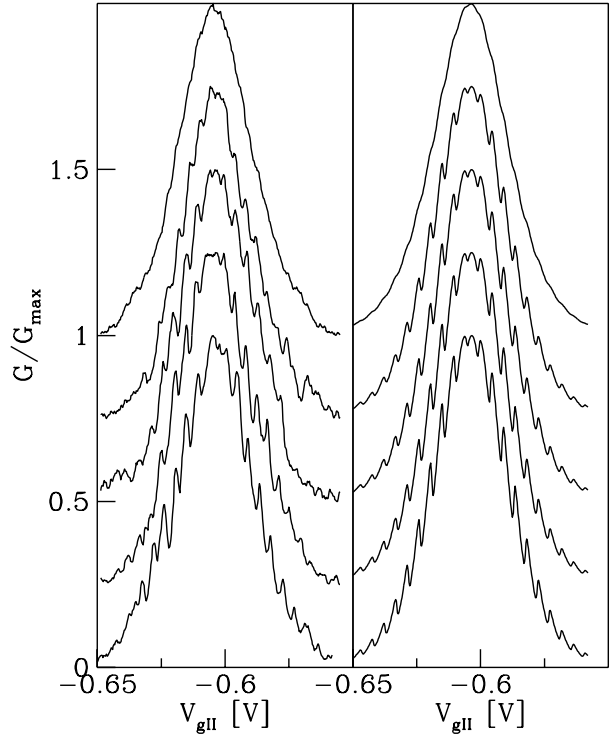}
   }
  \hss}
}
\caption{\label{fig:flucn2}
Traces of $G$ versus $V_{\rm{gII}}$ in an
electrometer experiment, where the conductance of barriers BC and DE is varied.
An offset of 0.25 $\times G/G_{\rm{max}}$ is used between consecutive curves.
Left panel: Experimental dat
a, where from top to bottom
$g_2,g_3 \approx $ 1.3, 0.65, 0.43, 0.14 
and 0.05 $\times e^2/h$.
In all traces $g_1,g_4,g_5 
\approx 0.05 \times e^2/h$. Right panel:
The results of our model calculations.}
\end{figure}

We will now discuss an experiment that measures the scaling behavior of 
$U_{2}$ with the conductance of barriers $g_2$ and $g_3$.
In the left panel of Fig.\ \ref{fig:flucn2} 
we plot $G$ versus $V_{\rm{gII}}$ in a series of measurements 
where the barriers between dot 2 and the wide 2DEG are gradually adjusted from 
the metallic to the tunneling regime. From top to bottom we have 
$g_2, g_3 \approx $ 1.3, 0.65, 0.43, 0.14 and 0.05 
$\times e^2/h$, 
respectively, while in all traces
$g_1,g_4,g_5 \approx 0.05 \times e^2/h$, 
so that dot 1 is always fully in the Coulomb blockade regime. 
One clearly observes the sawtooth structure on the dot-1 Coulomb 
oscillation due to the electrometer effect. In addition, one finds that 
for increasing conductances $g_2$ and $g_3$
the sawtooth feature is much less pronounced.

The right panel of Fig.\ \ref{fig:flucn2} 
shows lineshapes calculated from Eqs.
(\ref{Grenor}), (\ref{Erenor1}) and (\ref{scafor}). In order to obtain
a consistent set of fits, we first determine
the parameters  introduced in Sec.\ \ref{sec:model} ($U_{i}= 0.13$ meV, 
$U_{12} =$ 0.009 meV, $a_{ii} =-0.20$, and $a_{12}=-3.12$) 
from a fit of Eqs. (\ref{G}) and (\ref{Edelta}) to the
bottom trace of the left panel of Fig.\ \ref{fig:flucn2}
---, which is the same as the experimental (dotted) trace in the bottom
panel of Fig.\ \ref{fig:elmeter} --- 
where both dots are fully in the tunneling regime. The upper
curves are then obtained from Eqs. (\ref{Grenor}) and (\ref{Erenor1}), 
keeping the same values for $U_1$, $U_{12}$, and $a_{ij}$ while
varying $U_2^*$ with $g_2,g_3$ according to 
Eq.\ (\ref{scafor}) with $N_c=2$. 
This is because there are two contacts that allow tunneling.
The results of the model calculation is shown in the right hand
panel of Fig.\ \ref{fig:flucn2} and we see that very nice agreement
between theory and experiments is found.

\subsection{Determination of the effective charging energy}
\label{sec:fitting}

\begin{figure}
\vbox to 11cm {\vss\hbox to 11cm
 {\hss\
   {\includegraphics{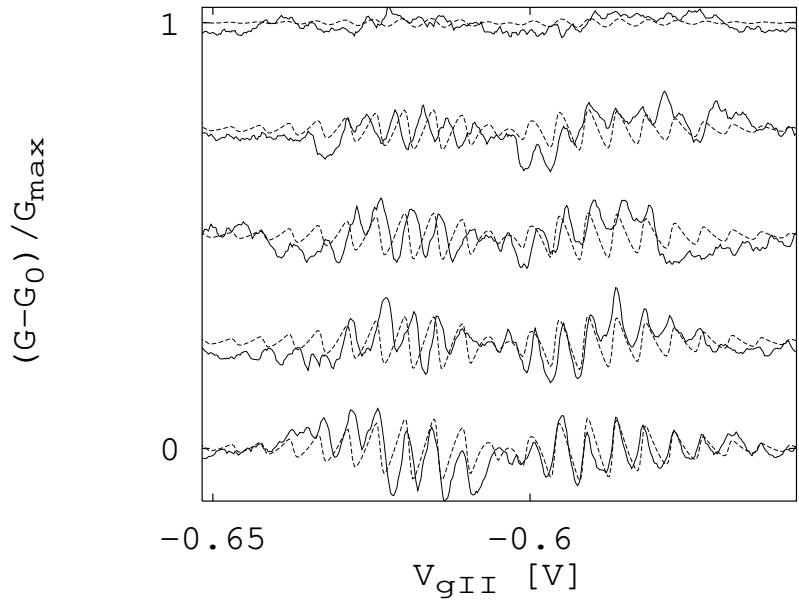}
   }
  \hss}
}
\caption{\label{fig:fit}
Traces of $G$ versus $V_{\rm{gII}}$ for the same data as in Fig.\
\protect\ref{fig:flucn2} but with the conductance corresponding to no
charging of dot 2 subtracted. See also Fig.\
\protect\ref{fig:elmeter}. Curves of this type is used to determine
the best fit for the renormalized charging energy $U_2^*$ and the
figure shows the best fit obtained.}
\end{figure}

\begin{figure}
\vbox to 6cm {\vss\hbox to 6cm
 {\hss\
   {\includegraphics{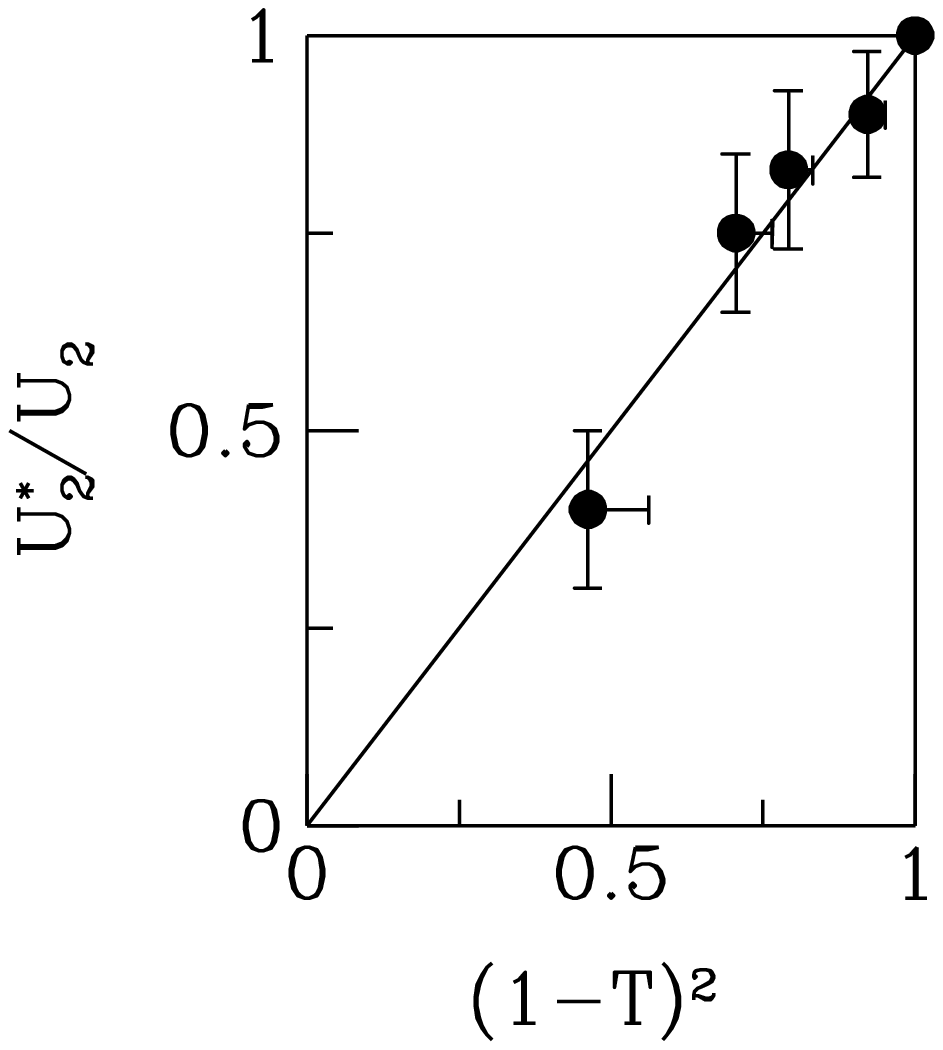}
   }
  \hss}
}
\caption{Plot of the ratio $U^*_2/U_2$ between "renormalized" and 
bare charging energy of dot 2, 
versus $(1-T)^2$, where $T$ is the transmission of barriers 
BC and DE that control the coupling between dot 2 and the external leads.
The linear dependence found
in this plot support the validity of a scaling law of the type in
Eq.\ (\protect\ref{scafor}).  
}
\label{fig:scaling}
\end{figure}

In order to determine the renormalized charging within our model
more accurately, we present in this section a fitting procedure
for finding $U_2^*$ according to Eqs. (\ref{Grenor}) and (\ref{Erenor1}). 
By subtracting the trace that is expected for a Coulomb island
without coupling to a dot 2 (the dashed curve in Fig.
\ref{fig:elmeter}), we obtain the traces shown in Fig.\ \ref{fig:fit}.
By doing the same for the model calculation we have producted the
dashed curves in Fig.\ \ref{fig:fit} by changing $U_2^*$ only. 
The optimum values of $U_2$ are found by comparing the average heights
of the peaks in subtracted curves thus producing a relation between the  
peak heights and $U_2^*$. This procedure allows us to determine 
the scaled charging energy of dot 2 fairly accurately. 
The results of our fitting procedure including estimates for the
uncertaincies are shown in Fig.\ \ref{fig:scaling}. The data is seen to 
be well described by the $(1-T)^2$ behavior predicted by Eq.\ (\ref{scafor}). 

\section{The one-to-two dots experiment, $N_c$=1}
\label{sec:flucndiffexp}

\begin{figure}
\vbox to 11cm {\vss\hbox to 11cm
 {\hss\
   {\includegraphics{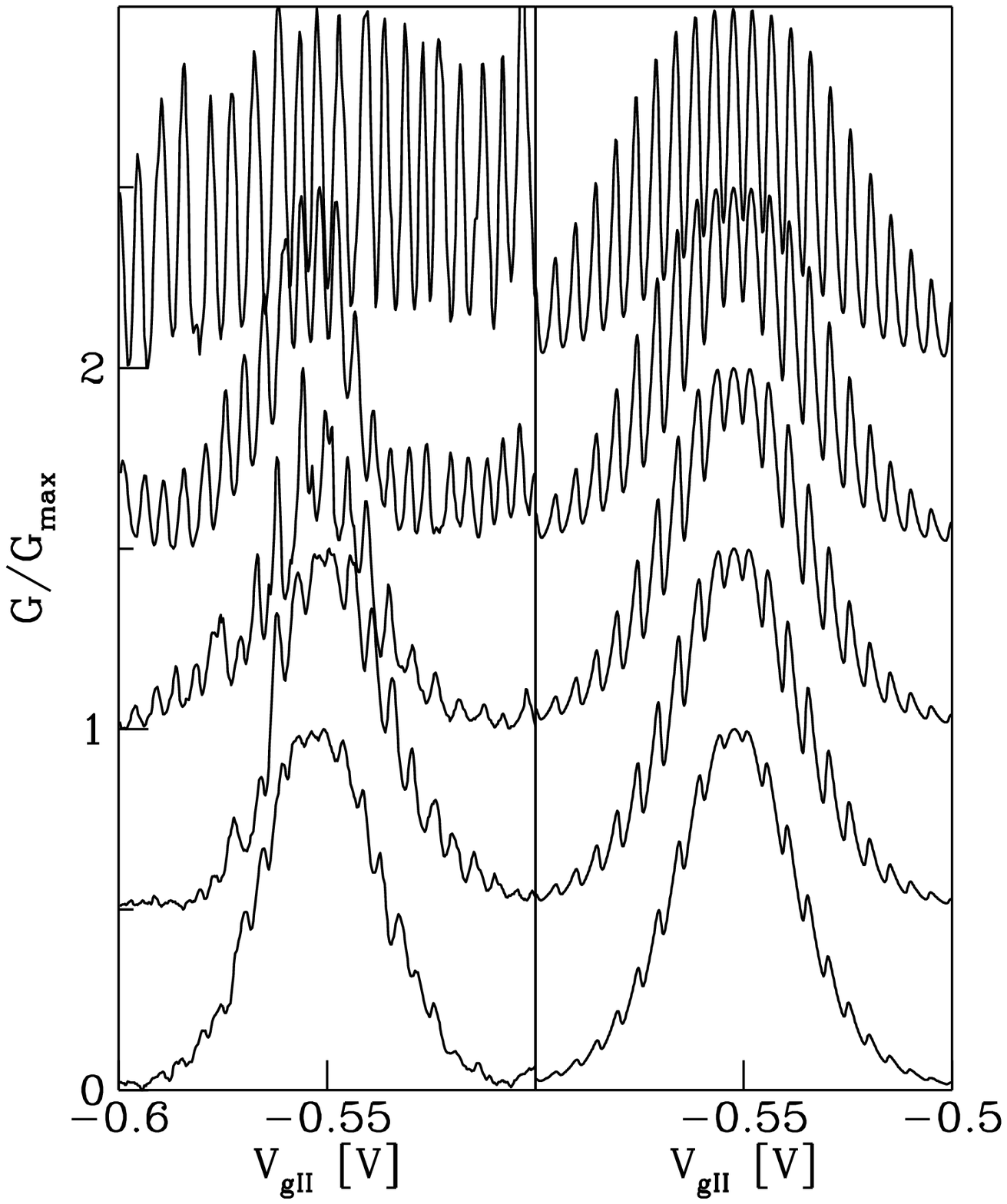}
   }
  \hss}
}
\caption{ $G$ versus $V_{\rm{gII}}$ in an experiment where 
$g_5$ is varied, 
so that the device changes from a single to a double quantum dot. 
An offset of 0.5 $\times G/G_{\rm{max}}$ is used between consecutive curves.
Left panel: Experimental traces, where, from top to bottom, $g_5=$
0.9, 0.5, 0.26, and 0.14 $e^2/h$, respectively. 
During this experiment all other tunnelbarriers are kept at a conductance of 
about 0.05 $e^2/h$.  
Right panel: 
Theoretical curves using Eq.\ (\protect\ref{Erenor2}) for $\delta E^*$.
}
\label{fig:onetotwo}
\end{figure}

Next we describe a second experiment where we study the 
transition from a single dot to a double dot Coulomb blockade device. 
This is done by adjusting the barrier BE, that connects the two dots
The data presented in the left panel of Fig.\ \ref{fig:elmeter} are plots of
$G (\equiv G_{14})$ versus $V_{\rm{gII}}$ for various adjustments of
the conductance $g_5$. 
Barriers AB, BC, DE, and EF are all adjusted
well into the tunneling regime, 
$g_1,g_2,g_3,g_4 \approx 0.05 \times e^2/h$. From top to bottom, 
$g_5$ varies from 0.9 to 0.14 $\times e^2/h$. 
Qualitatively one may interpret the data in this figure in a
straightforward manner: for $g_5 \approx 2e^2/h$
the device may be regarded as a single dot. One then expects the
rather regular, periodically spaced
Coulomb oscillations that are indeed observed in the top trace. 
On decreasing $g_5$, the electrons are confined more and more to 
either dot 1 or dot 2. As a result, 
a beating pattern evolves due to the differences in capacitance 
between gate II and dots 1 and 2. Finally,
for $g_5 \approx 0.01 \times e^2/h$ the device is in the
limit where dots 1 and 2 are two separate (but capacitatively coupled) 
quantum dots. 

It is clear that a quantitative discussion of the transition from 
one to dots must allow for fluctuations in the charge 
difference $(n_1-n_2)$ between the two dots. Such a model was
described above in Sec.\ \ref{sec:flucndiff}. The scaling of the 
charging energy that controls the charge difference between the two
dots, $U_{\Delta n}^*$, is assumed to be given by Eq.\ (\ref{scafor})
but now with $N_c=1$, since only one contact is connecting the dots.
Notice that in the experiment described previously, we had $N_c=2$.

The righthand panel of Fig.\  \ref{fig:onetotwo} shows the results
of the model calculation using Eqs. (\ref{Grenor}),(\ref{Erenor2}) and
(\ref{scafor}). We see that the data is reasonably well reproduced. 
However, we also note that our calculations display an increase in the width
of the dot 1 Coulomb resonance with increasing $g_5$, 
which is not observed experimentally. The reason for this discrepancy is
presently unclear.   

\section{Discussion and summary}
\label{sec:summa}

In conclusion, we have performed and analyzed experiments aimed at 
understanding the role of charge fluctuations in the transport properties
of quantum dots. We find that the dependence of the charging energy of a
quantum dot on the conductance of the point contact tunnel barriers 
can be well described using a scaling equation. We have speculated
that this behavior reflects an underlying one-dimensional physics.
The point contacts are one-dimensional in the sense that the
transmit only one channel. It has been argued that Coulomb blockade systems 
connected by quantum point contacts can be described in terms of
Luttinger liquid type models\cite{flen93capa,flen94capa,matv95} and we
view the present experimental results as support of this scenario.
Another recent experiment using a double dot structure\cite{waug95}
has also measured a change of the charging energy as a function of
the conductance of a point contact connecting the two dots. The
results of this investigation have also been interpreted in terms of
a Luttinger liquid type models\cite{matv:pre,gold:pre}. 
However, in that interpretation a more complete theoretical expression 
than our Eq.\ (\ref{scafor}) for the case $N_c=1$ has been used.

It would be useful to verify the validity of the scaling
equation for other power laws, which could be accomplished, 
e.g., by performing experiments in a high magnetic field, or by 
varying a different number of tunnelbarriers. 

{\em Acknowledgement} --- We thank M. Kemerink for performing part
of the experiments, and O. J. A. Buyk and 
M. A. A. Mabesoone for expert technical assistance.
The heterostructures were grown by 
C. T. Foxon at Philips Research Laboratories in Redhill (Surrey, UK)\@.

%\bibliographystyle{prsty}
%\bibliography{karsten}

\begin{thebibliography}{10}

\bibitem{likh:rev}
D.~V. Averin and K.~K. Likharev,  in {\em Mesoscopic Phenomena in solids},
  edited by B. L.Al'tshuler, P.~A. Lee, and R.~A. Webb (Elsevier, Amsterdam,
  1990).

\bibitem{metal:exp}
L. J. Geerligs, V. F. Anderegg, and C. A. van der Jeugd, Europhys. Lett. {\bf
  10}, 79 (1989); A. N. Cleland, J. M. Schmidt, and J. Clarke, Phys. Rev.
  Lett., {\bf 64}, 1565 (1990).

\bibitem{metal:theory}
G. Sch{\"{o}}n and A.~D. Zaikin, Phys. Rep. {\bf 198}, 237 (1990); A.~A.
  Odintsov, Zh. Eksp. Teor. Fiz. {\bf 94}, 312 (1988) [Sov. Phys. JETP {\bf
  67}, 1265 (1988)]; S.~V. Panyukov and A.~D. Zaikin, J. Low Temp. Phys. {\bf
  73}, 1 (1988); W. Zwerger and M. Scharpf, Z. Phys. B -- Condensed Matter {\bf
  85}, 421 (1991); K. Flensberg and M. Jonson, Phys. Rev. B {\bf 43}, 7586
  (1991); A.~D. Zaikin and S.~V. Panyukov, Physics Letters A {\bf 183}, 115
  (1993); H. Sch{\"o}eller and G. Sch{\"o}n, Physica B {\bf 203}, 423 (1994);
  G. Falci, J. Heins, G. Sch{\"o}n, and G. T. Zimanyi, Physica B {\bf 203}, 409
  (1994); H. Grabert, Phys. Rev. B {\bf 50}, 17364 (1994).

\bibitem{kouw91}
L.~P. Kouwenhoven {\it et~al.}, Z. Phys. B -- Condensed Matter {\bf 85},  367
  (1991).

\bibitem{foxm93}
E.~B. Foxman {\it et~al.}, Phys. Rev. B {\bf 46},  10020  (1993).

\bibitem{flen93capa}
K. Flensberg, Phys. Rev. B {\bf 48},  11156  (1993).

\bibitem{flen94capa}
K. Flensberg, Physica B {\bf 203},  432  (1994).

\bibitem{matv95}
K.~A. Matveev, Phys. Rev. B {\bf 51},  1743  (1995).

\bibitem{furu95}
A. Furusaki and K.~A. Matveev, Phys. Rev. Lett. {\bf 75},  709  (1995).

\bibitem{kane92fish}
C. L. Kane and M. P. A. Fisher, Phys. Rev. Lett. {\bf 68}, 1220 (1992); Phys.
  Rev. B {\bf 46}, 15233 (1992).

\bibitem{matv93glaz}
K. A. Matveev and L. I. Glazman, Phys. Rev. Lett {\bf 70}, 990 (1993); Physica
  B {\bf 189}, 266 (1993).

\bibitem{alei95glaz2}
I.L Aleiner and L.I. Glazman, Phys. Rev. B {\bf 52},  11296  (1995).

\bibitem{wen90}
X.G. Wen, Phys.\ Rev.\ Lett.\ {\bf 64}, 2206 (1990); Phys.\ Rev. B {\bf 41},
  12838 (1990): ibid {\bf 44} 5708 (1991).

\bibitem{mole95}
L.~W. Molenkamp, K. Flensberg, and M. Kemerink, Phys.\ Rev.\ Lett. {\bf 75},
  4282 (1995).

\bibitem{waug95}
F.~R. Waugh {\it et~al.}, Phys. Rev. Lett. {\bf 75},  705  (1995).

\bibitem{matv:pre}
K. A. Matveev, L. I. Glazman, and H. U. Baranger, cond-mat/9504099.

\bibitem{gold:pre}
J. M. Golden and B. I. Halperin, cond-mat/9505007.

\bibitem{lafa91}
P. Lafarge {\it et~al.}, Z. Phys. B {\bf 85}, 327 (1991); T. A. Fulton, P.L.
  Gammel, and L. N. Dunkleberger, Phys.\ Rev.\ Lett.\ {\bf 67}, 3148 (1991).

\bibitem{ruzi92}
I. Ruzin, V. Chandrasekhar, E. Levin, and L. Glazman, Phys. Rev. B {\bf 45},
  13469  (1992).

\bibitem{kuli75}
I. Kulik and R. Shehkter, Sov. Phys. JETP {\bf 41},  308  (1975).

\bibitem{glaz89}
L.~I. Glazman and R. Shehkter, J. Phys. Condens. Matter {\bf 1},  5811  (1989).

\bibitem{been91}
C. Beenakker, Phys. Rev. B {\bf 44},  1646  (1991).

\end{thebibliography}

\end{document}